%% file: main.tex
\documentclass[11pt]{article}
\usepackage{UF_FRED_paper_style}

\usepackage{lipsum}  
\usepackage{siunitx}
\usepackage{lineno,hyperref}
\usepackage[utf8]{inputenc}
\usepackage{geometry}
\usepackage{verbatim}
\usepackage{color}
\usepackage{natbib}
\usepackage{booktabs}
\usepackage{multirow}
\newcommand{\tabitem}{~~\llap{\textbullet}~~}

\usepackage{graphicx,amssymb,amsmath,subcaption}
\usepackage{diagbox}

\usepackage{etoolbox}

\title{Forex Trading Volatility Prediction using Neural Network Models
}

\author[1]{Shujian Liao\thanks{Email: shujian.liao.18@ucl.ac.uk.}}
\author[ ]{Jian Chen \thanks{Email: jian.j.chen@gmail.com.}}
\author[1,2]{ Hao Ni\thanks{Correspondence: h.ni@ucl.ac.uk.\href{https://orcid.org/0000-0001-5485-4376}{\includegraphics[scale=0.06]{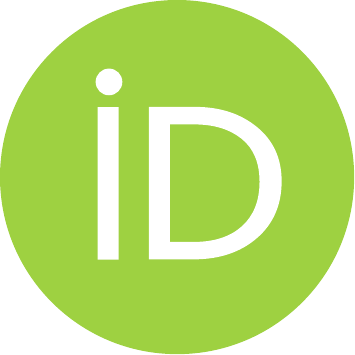}} ORCID: 0000-0001-5485-4376.} }

\affil[1]{Department of Mathematics, University College London}
\affil[2]{The Alan Turing Institute}

\date{\today}

\begin{document}
{\setstretch{.8}
\maketitle
\begin{abstract}

In this paper, we investigate the problem of predicting the future volatility of Forex currency pairs using the deep learning techniques. We show step-by-step how to construct the deep-learning network by the guidance of the empirical patterns of the intraday volatility. The numerical results show that the multiscale Long Short-Term Memory (LSTM) model with the input of multi currency pairs consistently achieves the state-of-the-art accuracy compared with both the conventional baselines, i.e. autoregressive and GARCH model, and the other deep learning models.

\noindent
\\
\textit{\textbf{Declaration of interests: }%
None.} 

\end{abstract}
}
\section*{Acknowledgements}
HN is supported by the EPSRC under the programme grant EP/S026347/1 and the Alan Turing Institute under the EPSRC grant EP/N510129/1. SL is supported by the UCL PhD Teaching Assistantship and Dean's Prize of Department of mathematics.
\newpage


\section{Introduction}
\subsection{Background and motivation}
 With the blossom of the real-world applications of Machine Learning (ML) over the last decade, it has become increasingly popular to apply ML in the financial industry~\cite{ftsreview}. Despite the enthusiasm, its success is arguably less prominent compared to other fields especially in trading. In this article, we explore the possibility of using deep learning methods~\cite{deeplearning} to solve a real-world problem closely related to trading applications, namely intraday volatility forecasting. In particular, we would like to demonstrate the power of an intuition-driven approach to use neural networks to model financial and economic phenomena. Our study shows that the inherent modular nature of deep neutral networks combined with the divide-and-conquer approach led by domain knowledge provides great potentials of building end-to-end solutions to solve the trading problems in the real life.

 Volatility modelling is one of the fundamental problems in the financial world. It is not only a crucial part for risk evaluations, but also a building block of trading processes. For example, option traders or investors' positions often reflect their views on the difference between the realised and implied volatility. High frequency trading algorithms, either for speculative or market-making purposes, need to predict intraday volatility to make real-time trading decisions. However, despite the widely accepted concept of volatility, the practitioner often utilises ranges, i.e. the differences between the maximum and minimum of a quantity within a specified time period, as a proxy of volatility in real trading. While volatility is a mathematical concept based on a particular model assumption, e.g. Geometric Brownian Motion, as discussed in~\cite{range-based_volatility}, range is closely linked with volatility, but it has a major advantage: it is an observable quantity directly linked to the maximum P\&L of trading positions. Its practicality makes it more popular with traders in their decision making processes. As a result, we study ranges as opposed to volatility in this article. We use spot foreign exchange (FX) data to forecast intraday ranges. However, the methodology can be applied on other asset classes.

The application of deep learning in volatility forecasting has been an active area in finance, where deep neural network (DNN) and recurrent neural network (RNN) are two popular methods. On the one hand, DNN is a common network architecture, which takes multivariate features as inputs and feeds them into multiple fully connected layers. The combination of the DNN models with the fundamental features achieves the superior performance in terms of the predictive modelling, for example, prediction of the cross-sectional stock markets in Japan and S$\&$P 500 index (e.g. \cite{dlcs_abe} and \cite{dlpar_feng}).  %
On the other hand, due to the strength of capturing the temporal information of data, RNNs or its variants (e.g. Long short-term memory (LSTM), Gated recurrent unit (GRU), etc) have achieved remarkable results for sequential data modelling, such as speech recognition, language forecasting, etc.~\cite{Sundermeyer2012LSTMNN, Speechrecognition}. For the volatility forecast task, \cite{xiong2016deep} shows the RNN type model coupled with price data of S$\&$P 500 and Google trends data outperform the Ridge/Lasso regression and the GARCH model. It is followed by \cite{csi2018}, in which the authors considered a single layer LSTM with the input including key words searching volume from Baidu and historical price data, which beats the GARCH model on CSI300 volatility dataset.

 \subsection{Patterns in the intraday ranges}
 The predictability of any quantity originates from its repetitive patterns often as a consequence of physical mechanisms behind the scene. FX ranges exhibit strong patterns, which usually have economic reasons and are well-understood by traders. These patterns and the domain knowledge of the causes of the patterns lay the foundation of our modelling process. Our main goal is to capture these economic linked intuitions in our prediction model.

\begin{figure}[h]

    \includegraphics[width= 1\textwidth]{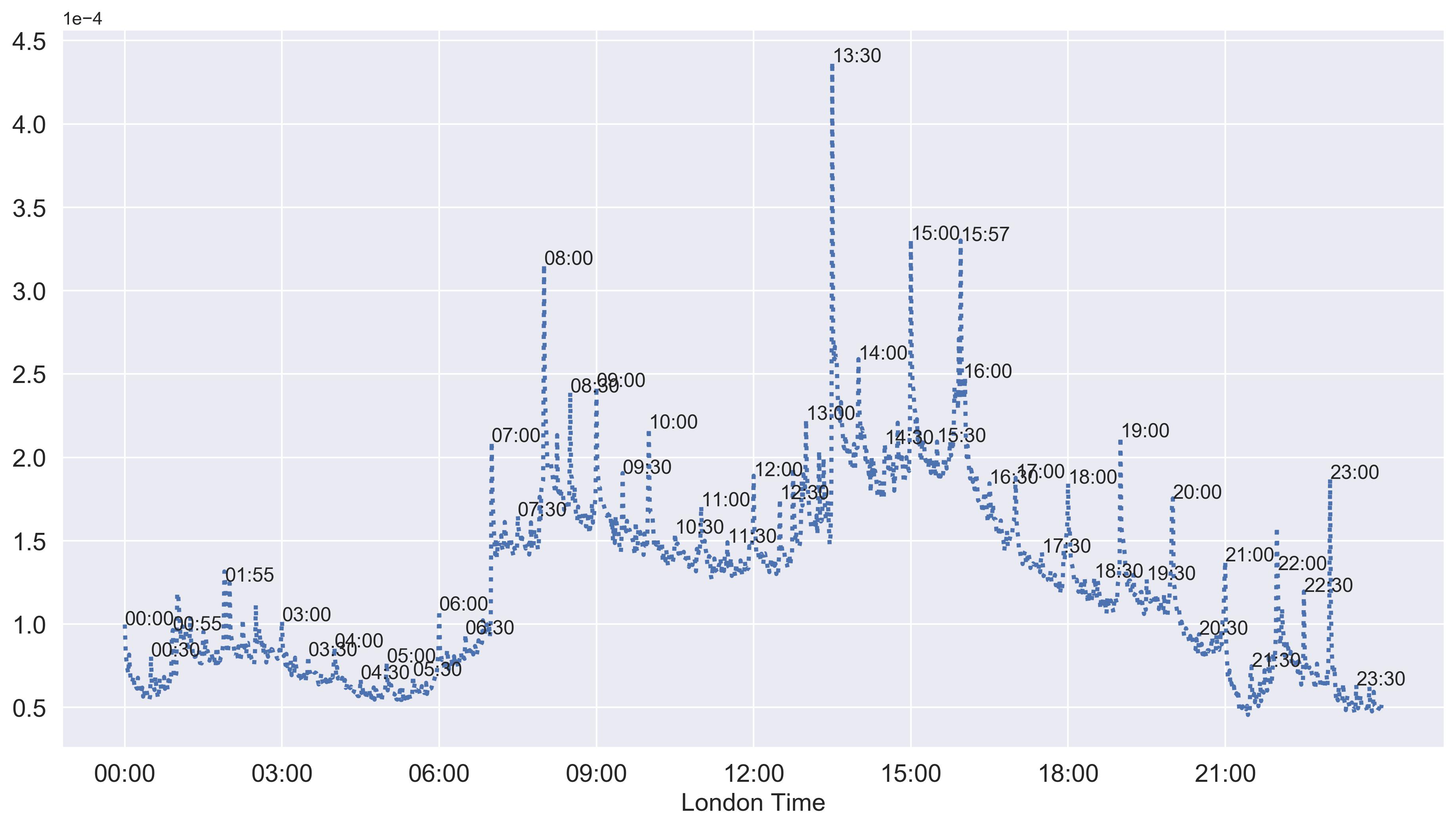}

    \caption{The daily log-range average for each minute. (Color)}
    \label{eurusd_pattern_spikes}
\end{figure}

 According to~\cite{range-based_volatility}, consider a finite interval of length $\tau$ and let $p_t$ be the price of currency at time $t$, then the log range is defined as
 \begin{equation*}
     \log(\sup_{0\leq t\leq \tau} p_t)-\log(\inf_{0\leq t\leq \tau} p_t).
 \end{equation*}
 Figure~\ref{eurusd_pattern_spikes} shows the average of log range for each minute using intraday EURUSD spot FX data from 01/01/2018 to 31/12/2019.

It exhibits interesting time-related patterns including:
\begin{itemize}
    \item The ranges tend to be bigger in London trading hours between 7am and 6pm London time corresponding to more frequent trading activities.
    \item Two humps around 7am and 11am London time corresponding London and New York market open.
    \item Various spikes in the ranges correspond to different economic data releases (e.g. Non-farm payroll at 1:30pm) and fixing trades (e.g. WMR fixing at 4pm).
\end{itemize}
Apart from the observations above, it is also well-known that FX market is heavily influenced by macro economic news and events. When market absorbs certain unexpected news, the stimulated moves often persists for a while. Because of the intrinsic correlations between different currency pairs, the level of activeness of the whole market are also inter-linked across all major currency pairs. As a result, we often observe the volatility clustering and cross-currency correlations.

The patterns discussed above can be summarised into 3 main categories:
\begin{itemize}
    \item time-based seasonality, e.g. trading hour related patterns, week-day patterns, etc.
    \item intraday auto-correlations, i.e. volatility clustering.
    \item cross currency pair correlations.
\end{itemize}

These patterns are backed by economic reasons and hence provide predictability. However, it is clearly a non-trivial problem to capture all these patterns in a single model.

 \subsection{The road map for deep learning of FX ranges}
Despite the complexity of the task, the fundamental problem is to build a function, which can be updated in real time to predict future ranges. In other words, the goal is to build a model to generate range functions in real time. As we can see from~\cite{range-based_volatility}, the function form is "unconventional" in the sense that it can have spikes and abrupt changes seemly arbitrarily. This presents the main challenge to traditional modelling approaches, which require building multiple separate models to capture individual patterns introducing model inconsistency. However, neural network, as a non-parametric method, is a perfect candidate for solving our problem because it can provide the flexibility required to produce the highly "customised" function form defined by FX market. Because of the inherent modular natural of neural networks, i.e. the combination of neural networks is still a neural network, it also allows us to model the 3 main patterns listed above individually using a divide-and-conquer approach. Once these patterns are separately learned by neural networks with different structures, they can then be naturally combined into a single architecture to capture all the features. From the modelling perspective, this is not only satisfying, but also provides a generic framework to build end-to-end solutions for much more complicated problems. As an example, if we want to feed the predicted ranges into a trading algorithm, it can be further built on top of the neural works for ranges.

\subsection{The structure of the paper}
Following the road map described in the last section, this paper is organised as follows: in section \ref{Section_Methodology}, we start with the empirical analysis on the intraday volatility patterns and then illustrate how we incorporate domain knowledge in designing our models for range forecast. We conduct the numerical experiment and provide the comparative results in section \ref{Section_Result}. Section \ref{Section_Conclusion} concludes the paper. For readers who are not familiar with neural network, please refer to \ref{section_appendix} for the preliminary of neural networks.

\section{Data description}\label{Section_Data_description}
In this section, we describe the FX dataset for the task of intraday volatility prediction and illustrate its empirical patterns, which are major consideration for our neural network model design.

We use the minutely price information of four Forex cross pairs (i.e. EURUSD, EURSEK, USDJPY and USDMXN) as the example data, which is retrieved from Histdata\footnote{https://www.histdata.com/download-free-forex-data/?/excel/1-minute-bar-quotes}. It includes time stamps, open price, high price, low price and close price. The time period is from 2018-01-01 to 2019-12-31, and the daily trading period is of $T=1440$ minutes. Table~\ref{rawdata_eurusd} is a snapshot of the price data of EURUSD.
\begin{table}[htbp]
    \centering
    \begin{tabular}{cccccc}
    \toprule
        \textbf{Date} & \textbf{Time} & \textbf{Open} & \textbf{High} & \textbf{Low} & \textbf{Close}\\
        \midrule
       \multicolumn{1}{c}{01/01/2018}  & \multicolumn{1}{c}{22:00} & \multicolumn{1}{c}{1.20037} & \multicolumn{1}{c}{1.20100} & \multicolumn{1}{c}{1.20037} & \multicolumn{1}{c}{1.20100}\\
       \multicolumn{1}{c}{01/01/2018}  & \multicolumn{1}{c}{22:01} & \multicolumn{1}{c}{1.20083} & \multicolumn{1}{c}{1.20095} & \multicolumn{1}{c}{1.20017} & \multicolumn{1}{c}{1.20030}\\
       \multicolumn{1}{c}{01/01/2018}  & \multicolumn{1}{c}{22:02} & \multicolumn{1}{c}{1.20035} & \multicolumn{1}{c}{1.20043} & \multicolumn{1}{c}{1.20035} & \multicolumn{1}{c}{1.20043}\\
       \multicolumn{6}{c}{\vdots}\\
       \multicolumn{1}{c}{12/31/2019}  & \multicolumn{1}{c}{21:57} & \multicolumn{1}{c}{1.12115} & \multicolumn{1}{c}{1.12115} & \multicolumn{1}{c}{1.12105} & \multicolumn{1}{c}{1.12105}\\
       \multicolumn{1}{c}{12/31/2019}  & \multicolumn{1}{c}{21:58} & \multicolumn{1}{c}{1.12105} & \multicolumn{1}{c}{1.12110} & \multicolumn{1}{c}{1.12099} & \multicolumn{1}{c}{1.12099}\\
       \multicolumn{1}{c}{12/31/2019}  & \multicolumn{1}{c}{21:59} & \multicolumn{1}{c}{1.12099} & \multicolumn{1}{c}{1.12115} & \multicolumn{1}{c}{1.12076} & \multicolumn{1}{c}{1.12076}\\
       \bottomrule
    \end{tabular}
    \caption{The minutely price data of EURUSD pair from 2018-01-01 to 2019-12-31.}
    \label{rawdata_eurusd}
\end{table}

\subsection{Time-based Seasonality}\label{subsec: seasonality}

\subsubsection{Weekly Seasonality}
\begin{figure}
        \centering
        \includegraphics[width= 1\textwidth]{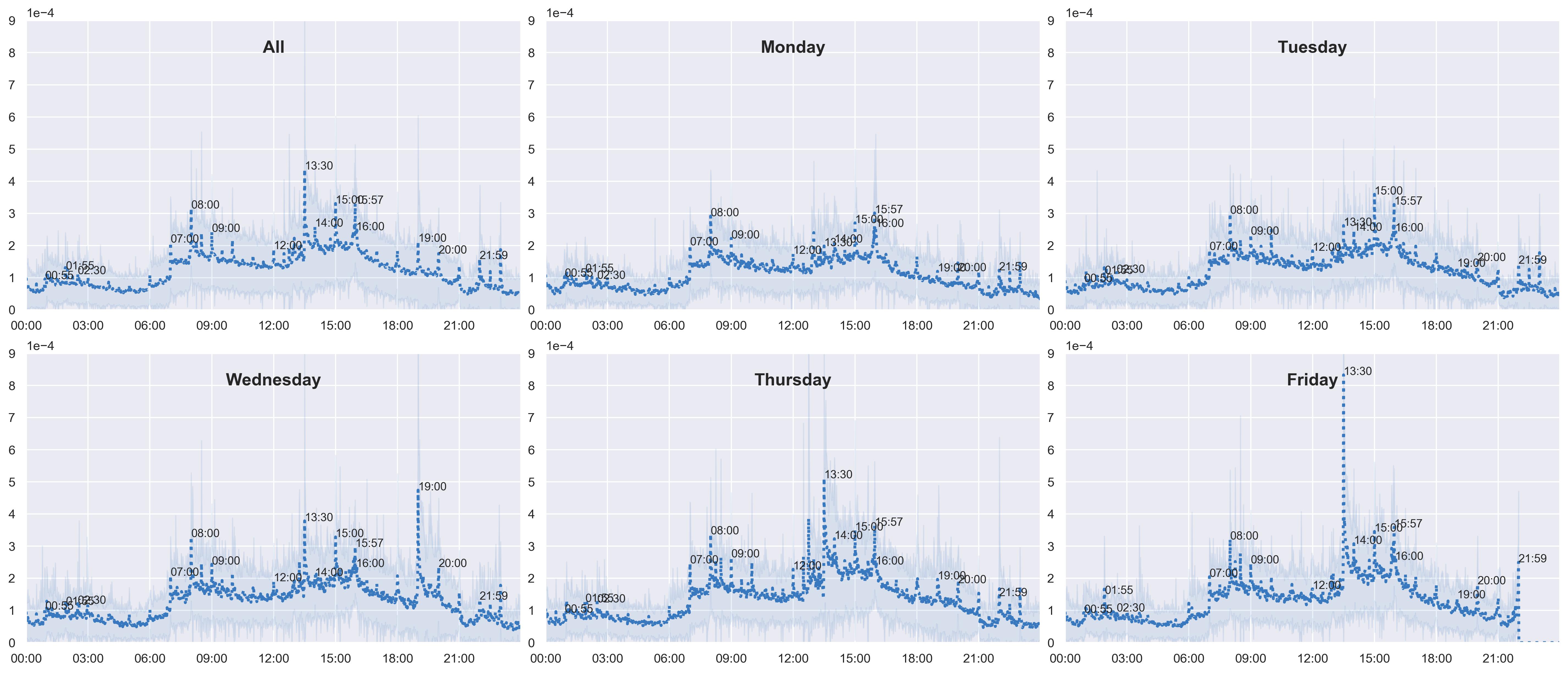}
        \caption{Weekly seasonality of the trading volatility. (Color)}
        \label{seasonality}
    \end{figure}

Figure~\ref{seasonality} plots the average log range for each weekday. It shows the significant difference between different weekday and weekly seasonality in trading volatility in a clear way. According to Figure \ref{seasonality}, we observe:
\begin{itemize}
        \item The trading volatility on Monday and Tuesday are slightly lower than other days on average.
        \item Some spikes are only obvious on certain weekdays. For example, the 13:30 spike mainly appears on Friday, Wednesday and Thursday, especially on Friday when Non-farm payroll is released. The 19:00 spike is only significant on Wednesday, when the FOMC minutes happens.
        \item Fixing spikes at 00:55, 01:55, 07:00, 08:00, 09:00 and 15:57 appear every day and their volatility are relatively even across the whole week.
    \end{itemize}

The intraday volatility patterns can help traders to better predict the intraday trading volatility and hence select a better time to execute their trades.
\subsubsection{The Spikes}
In Figure~\ref{eurusd_pattern_spikes}, we plot the daily log range average for each minute. Most of these spikes or sudden increase of volatility can be associated with, for example, economic events:
\begin{itemize}
    \item Spikes typically appear at minute 0, 30 and 45 of each hour. These spikes normally correspond to economic data releases at 00, 30 and 45.
    \item Spikes also appear at 00:55, 01:55, and 16:00. These correspond to the Tokyo and WM fixing times respectively.
    \item Trading volatility are significantly higher during the London trading hours.
\end{itemize}

\subsection{Intraday and Interday Auto-correlation}\label{intra_inter_autocorr}
\begin{figure}[!ht]
    \begin{subfigure}{0.5\linewidth}
    \centering
   \includegraphics[width=1\textwidth]{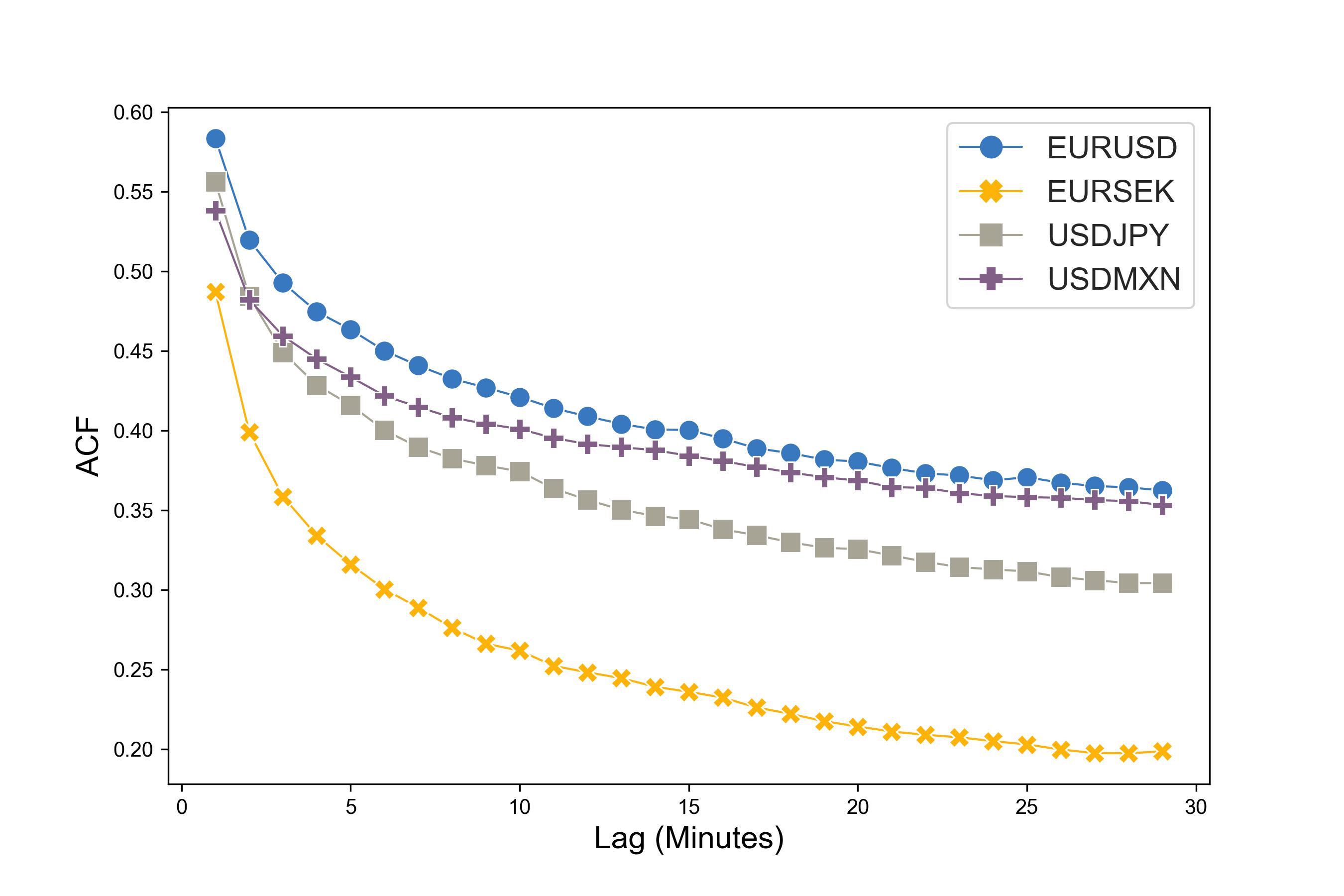}
   \end{subfigure}
       \quad
           \begin{subfigure}{0.5\linewidth}
           \centering
   \includegraphics[width=1\textwidth]{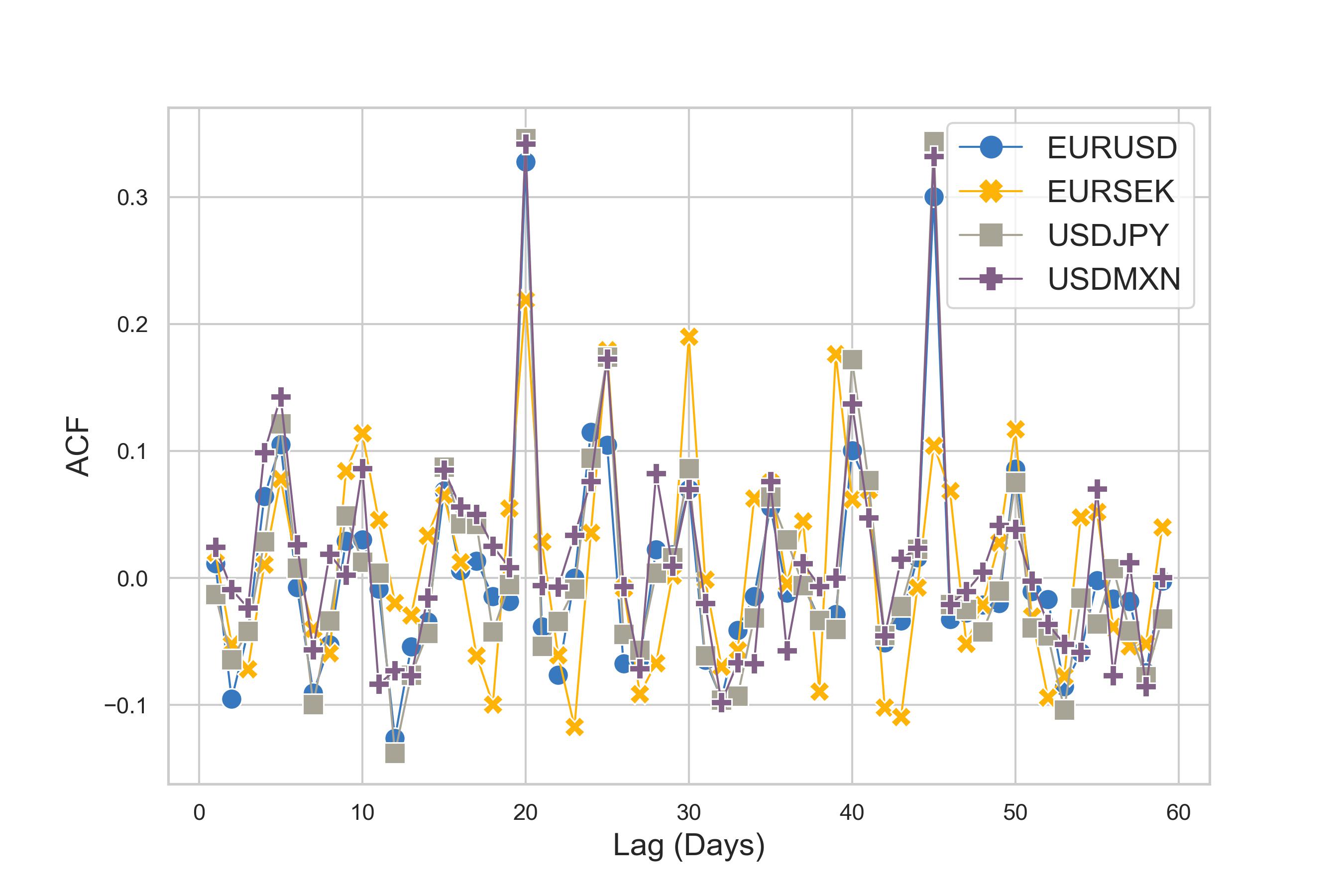}
   \end{subfigure}
\caption{(Left) The intraday autocorrelation of log range; (Right) The interday autocorrelation of log range at 13:30 London time.(Color)}\label{acf_plot}
\end{figure}

The intraday auto-correlations reflect the latency in trading volatility, i.e. large changes usually immediately followed by large changes. The inter-day auto-correlations show the daily seasonality of volatility.  Shown in Figure~\ref{acf_plot}, the intraday auto-correlation is relatively high for the first $1$ lag minutes and decays sharply, which means that the investment inertia last a minute and then disappears. The volatility clustering can be observed from the serial positive intraday auto-correlation.

In terms of the interday volatility, we investigate the auto-correlation of the most significant spike, i.e. at 13:30. When the lag value equals 20 weekdays, the auto-correlation is the highest. This is expected, as 13:30 corresponds to the Nonfarm payroll, which is released once a month.
\subsection{Cross currency pair correlation}\label{subsec:cross_correlation}
\begin{figure}[ht]
    \centering
    \includegraphics[width= 1\textwidth]{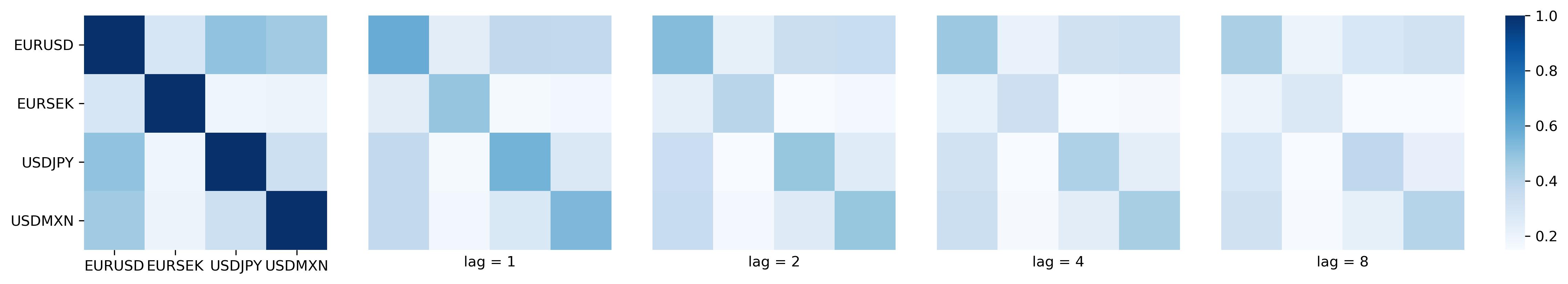}
    \caption{The (time lagged) correlations plot of cross currency with respect to the number of lags, where the lag values are 0, 1, 2, 4, 8 minutes. (Color)}
    \label{cc_plot}
\end{figure}

Illustrated in Figure~\ref{cc_plot}, the (time lagged) correlation matrix plot of four cross currency pairs shows that the pairs sharing same base or quote currencies have much higher correlation, i.e. USDJPY-EURUSD and USDMXN-EURUSD. In contrast, the pairs that have no common currencies are less correlated, i.e. USDMXN-EURSEK and USDJPY-EURSEK.


\section{Methodology}\label{Section_Methodology}

The main objective of our work is to build an effective neural network model, which has a built-in structure designed to capture the above-mentioned empirical patterns of FX data for the volatility prediction. In this section, we show step by step how to build a single neural network to capture all the desired empirical characteristic of volatility individual by (1) first designing each module to capture one of empirical patterns, e.g. the seasonality, time auto-correlation and cross currency pairs dependence; (2) then integrating those modules to construct a whole network architecture.

Let us introduce the set-up of our problem and the necessary notations for the ease of our discussion. Let $\mathcal{D}$ denote the total market days in the dataset, and $T$ denote the total minutes of each market day. Let $V_t^D\in \mathbb{R}$ denote the log range of a currency pair at time $t$ of day $D$, where $t\in\{1,\dots,T\}$ and $D\in\{1,\dots, \mathsf{D}\}$. We aim to predict the next minutely log range $V_{t+1}^{D}$ using the information up to time $t$ at day $D$, which includes all the previous log range data and its corresponding date and time.

We adopt the standard regression setting. (More details can be found in \ref{section_supervised_learning}.) Suppose we have $X_{t}^{D} \in \mathcal{X}$ to denote the explanatory variable to predict $V_{t+1}^{D}$. Let $\mathcal{D}$ denote the input-output pairs $\left (X_{t}^{D}, V_{t+1}^{D}\right)_{t, D}$. Let $f_{\Theta}: \mathcal{X} \rightarrow \mathbb{R}$ denote the model, which is fully characterized by the parameter set $\Theta$. We use the squared loss function to optimize the model parameters to fit the data $\mathcal{D}$, i.e.
\begin{eqnarray*}
\text{Loss}( \Theta \vert \mathcal{D} ) = \sum_{t, D} \left(V_{t+1}^{D} - f_{\Theta}(X_{t}^{D})\right)^{2}.
\end{eqnarray*}

In the following, we discuss the choices of input $X_{t}^{D}$ and model architecture $f_{\Theta}$, and show how to build a unified neural network to capture all the above-mentioned empirical patterns. The preliminary of the neural network, including the framework of the supervised learning, deep neural network (DNN) and Long short-term memory (LSTM) model, can be referred to \ref{section_appendix}.

\subsection{DNN and time-based patterns}\label{sectionDNN}
The trading volatility demonstrates the weekly and monthly seasonality shown in Figure \ref{seasonality} of section \ref{subsec: seasonality}. To capture the multi-scale seasonality, we consider the time stamp $t$, the day of week ($day\_of\_week_t^D$) and the month ($month_t^D$) as the input factors for volatility prediction. In addition, it is common for investors to mark their trading strategies to one fixing on the last business day of the month, which results in higher volatility of the month-end day compared with that of non month-end day. Therefore, we choose
$x_t^D = (t, day\_of\_week_t^D, month_t^D $ $,is\_month\_end_t^D) \in \mathbb{R}^4$ as the input factors. To start with, we use a multi-layer artificial neural network (DNN) as a model to learn the next minutely volatility. This method is denoted by Plain DNN.

\subsection{LSTM and auto-correlations}
The volatility usually exhibits the clustering behaviour. To capture the strong intraday (interday) auto-correlation of the volatility series, we use the lagged value of log range in the previous $p_t$ minutes (the log range at the same $t$ in the previous $p_d$ days) as the input $y_{t}^{D}$ ($z_{t}^{D}$) respectively, i.e.
\begin{eqnarray}
y_t^{D} &=& (V_{t-p_t}^{D},\dots,V_{t-1}^{D})\in \mathbb{R}^{p_t\times 1}\nonumber\\
z_t^{D} &=& (V_{t}^{D-p_d},\dots,V_{t}^{D-1})\in \mathbb{R}^{p_d\times 1}\label{rnn_inputs}
\end{eqnarray}
The LSTM model and its variants are well-known to have the strength in analysing sequential data. As $y_{t}^{D}$ or $z_{t}^{D}$ are time series, we choose to use the LSTM model which takes $y_{t}^{D}$ ($z_{t}^{D}$) as the input and output the predicted log-range, which we denote as $\text{LSTM}_t$ ($\text{LSTM}_D$) correspondingly. We call them the plain LSTM models.
\subsection{2-LSTM and multi-scale time dependence}
\begin{figure}
    \centering
    \includegraphics[width= 0.7\textwidth]{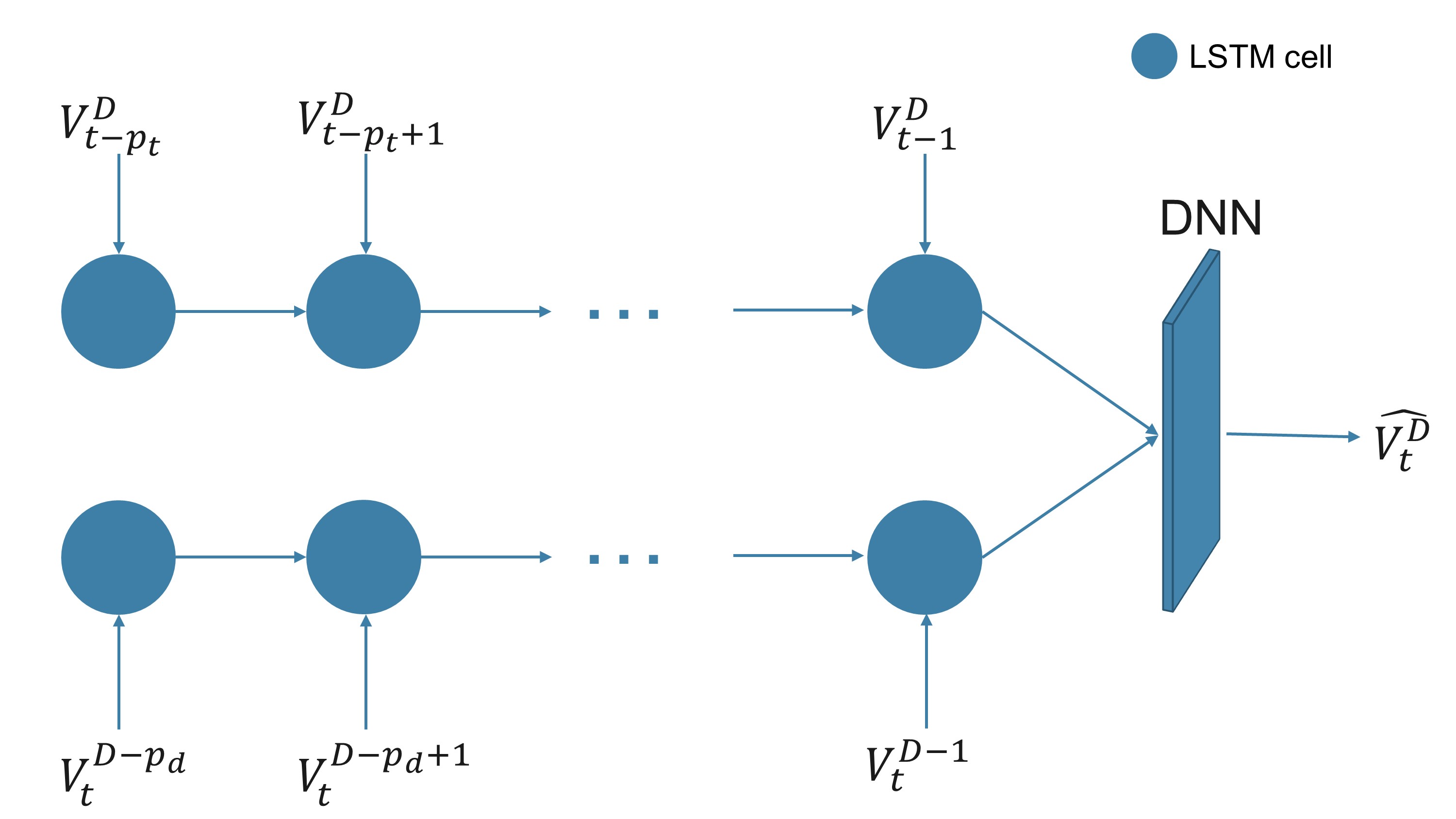}
    \caption{Illustration of 2-LSTM architecture. (Color) }
    \label{2lstm_flow}
\end{figure}

The above LSTM$_t$ or LSTM$_D$ can only model the dependence of the consecutive intraday volatility or interday volatility alone. To capture both seasonality of the volatility simultaneously, we propose the multi-scale LSTM model, namely 2-LSTM model. Let us consider $x_{t}^{D} = [y_{t}^{D}, z_{t}^{D}]$, where $[y_t^{D},z_{t}^{D}]$ is defined in~\eqref{rnn_inputs}. As shown in Figure~\ref{2lstm_flow}, we can construct a 2-LSTM model to forecast the volatility $V_t^D$ as follows: we first applies two LSTM models to $y_{t}^{D}$ and $z_{t}^{D}$ respectively, denoted by $\text{LSTM}(y_{t}^{D})$ and  $\text{LSTM}(z_{t}^{D})$; we then concatenate the outputs of those two LSTMs and apply the $l$-layer DNN to the concatenated output. Formally, 2-LSTM model can be expressed in the below formula:
\begin{equation}
    f_{\Theta}(x_{t}^{D})= \text{DNN}(\text{LSTM}(y_{t}^{D}),\text{LSTM}(z_{t}^{D})).    \label{2_rnn_model}
\end{equation}

\subsection{$p$-Pairs-learning 2-LSTM and cross currency pair correlations}
It is very challenging to learn the volatility of the non-liquid currency pairs due to infrequent trades and price changes. It is shown in section \ref{subsec:cross_correlation} that those currency pairs, which share same base, are often driven by common events, even if the liquidity of those pairs may be very different. It suggests that by coupling the non-liquid currency pairs information with that of other more liquid pairs with shared currency base, the volatility prediction may be further improved. Hence, we propose the $p$-pairs learning 2-LSTM model. We denote the volatility of $p$ currency pairs by $\textbf{V}_t^D = (V_{1,t}^D,\dots, V_{p,t}^D) \in \mathbb{R}^{p}$. Similar to~\eqref{rnn_inputs}, the input is defined as $\textbf{x}_{t}^{D} = [\textbf{y}_{t}^{D}, \textbf{z}_{t}^{D}]$, where
\begin{eqnarray}
\textbf{y}_t^{D} &=& (\textbf{V}_{t-p_t}^{D},\dots,\textbf{V}_{t-1}^{D})\in \mathbb{R}^{p_t\times p}\nonumber\\
\textbf{z}_t^{D} &=& (\textbf{V}_{t}^{D-p_d},\dots,\textbf{V}_{t}^{D-1})\in \mathbb{R}^{p_d\times p}\label{p_rnn_inputs}.
\end{eqnarray}
Then followed by~\eqref{2_rnn_model}, the $p$-Pairs-learning 2-LSTM model is fully constructed. The $p$-Pairs-learning model is a generalization of the -LSTM model as the 2-LSTM is the special case of $p$-Pairs learning for $p = 1$. When $p \geq 2$, the volatility forecast of each pair by the $p$-Pairs-learning can use the historical information of the other pairs shared with the same currency base, which better capture the cross currency pair correlation. To avoid the ambiguity of the $p$-Pairs-learning model with 2-LSTM, we assume $p \geq 2$ throughout the rest of the paper.

\section{Numerical Results}\label{Section_Result}

In this section, we compare the predictive performance of the deep-learning based models (i.e. Plain DNN, Plain LSTM, $2$-LSTM and $p$-Pairs learning) and benchmark them with two traditional autoregressive models (i.e. the autoregressive (AR($p$)) model and GARCH model). After the hyper-parameter tuning, we choose the optimal orders $p$ for EURUSD, EURSEK, USDJPY, and USDMXN are $1, 1, 2, 5$ respectively based on the results shown Figure~\ref{ar_order_select} of Appendix~\ref{hp_tune}. Following the work of volatility forecasting \cite{forecast_vol_THR},\cite{KIM_forecast_vol_lstm},\cite{forecast_vol_sgp_kuen},\cite{forecast_vol_garch_franses}, we specify $p, q = 1$ for the $GARCH(p,q)$, which is a widely-used model of generalized autoregressive models to capture volatility clustering.

\begin{figure}
        \centering
        \includegraphics[width= 1\textwidth]{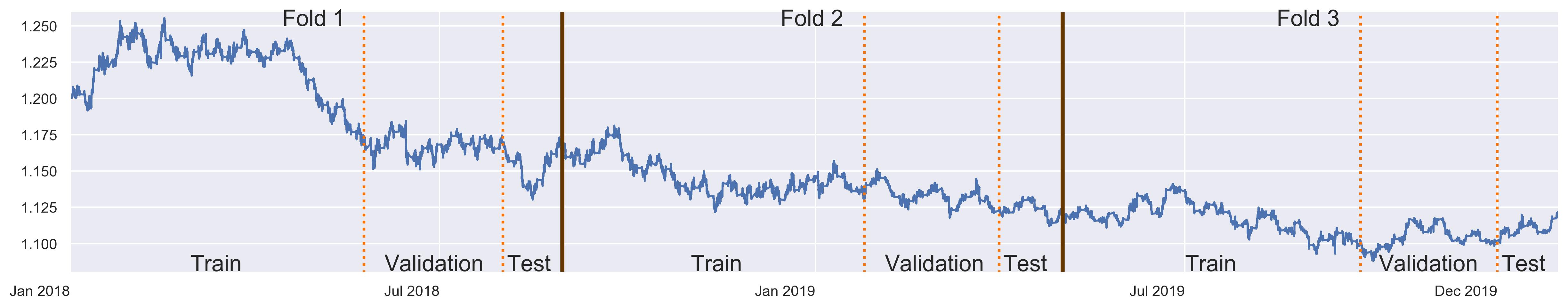}
        \caption{Illustration of the 3-fold chronological validation. (Color)}
        \label{cross_val}
    \end{figure}

We conduct 3-fold sequential blocked cross validation~\cite{ml_default_debt} in chronological order, and each fold is split into training, validation and testing sets with the corresponding percentage $0.6$, $0.3$ and $0.1$, as illustrated in Figure~\ref{cross_val}. We choose the mean squared error (MSE~\eqref{mse}) on the testing data sets as the test metric to assess the model performance. To further investigate whether the comparison between models performance is significant, we conduct a statistical test called Diebold-Mariano (DM) test~\cite{dmtest} between all the models in a pair-wised manner for all the currency pairs. We pre-process the log range data by the min-max normalization for all the model training.

\subsection{Network architecture}
We conduct the thorough hyper-parameter tuning on the proposed deep-learning based models. Interested readers can refer the implementation details to Appendix~\ref{hp_tune}. In the following numerical analysis, we specify the optimal model architecture as follows:

\begin{itemize}
\item Plain DNN: The optimal architecture are chosen as $6$ layers with $30$ neurons per layer.

\item Plain LSTM, 2-LSTM and $p-$pairs learning: For the LSTM based models, we set lag values $p_t = 20$ and $p_d = 20$. The number of hidden neurons is set to be $64$. The last DNN module $h_L$ has $L = 2$ with $32$ hidden neurons.

\end{itemize}

\subsection{Model comparison}

 \begin{table}[!ht]
     \centering

     \scalebox{0.8}{\begin{tabular}{l|l}
     \toprule
      \multicolumn{1}{c|}{\textbf{Advantages}} & \multicolumn{1}{c}{\textbf{Disadvantages}} \\
     \midrule
      \multicolumn{2}{c}{DNN}\\
     \midrule
     \tabitem Easy to create.& \tabitem Does not capture seasonality and spikes.  \\
     \tabitem Quick to train. & \tabitem Can not predict daily volatility.\\
     \tabitem Capture the general trend of average volatility.&\\
     \midrule
  \multicolumn{2}{c}{ LSTM}\\
  \midrule
      \tabitem Capture inter-day or intraday autocorrelation.&\tabitem Does not capture both autocorrelation.\\
     \tabitem Lag value can be adjusted for long or short term.&\tabitem Lagging in the prediction against target volatility\\
     \midrule

\multicolumn{2}{c}{ 2-LSTM}\\
\midrule
\tabitem Capture both inter/intra-day autocorrelation. &\tabitem Lagging in the prediction against target volatility.\\
     \tabitem Lag value can be adjusted for long or short term.&\\
     \midrule

\multicolumn{2}{c}{     $p$-pairs-learning 2-LSTM}\\\midrule
     \tabitem Capture both inter/intra-day autocorrelation.&\tabitem Lagging in the prediction against target volatility. \\
     \tabitem Capture cross currencies correlations.&\tabitem Further research on choosing currency pairs \\
     \tabitem Lag value can be adjusted for a either long or short term.& \\
     \bottomrule
     \end{tabular}}
     \caption{Summary of Volatility Forecasting Methods}\label{models_summary}
\end{table}

\begin{figure}
        \centering
        \includegraphics[width= 1\textwidth]{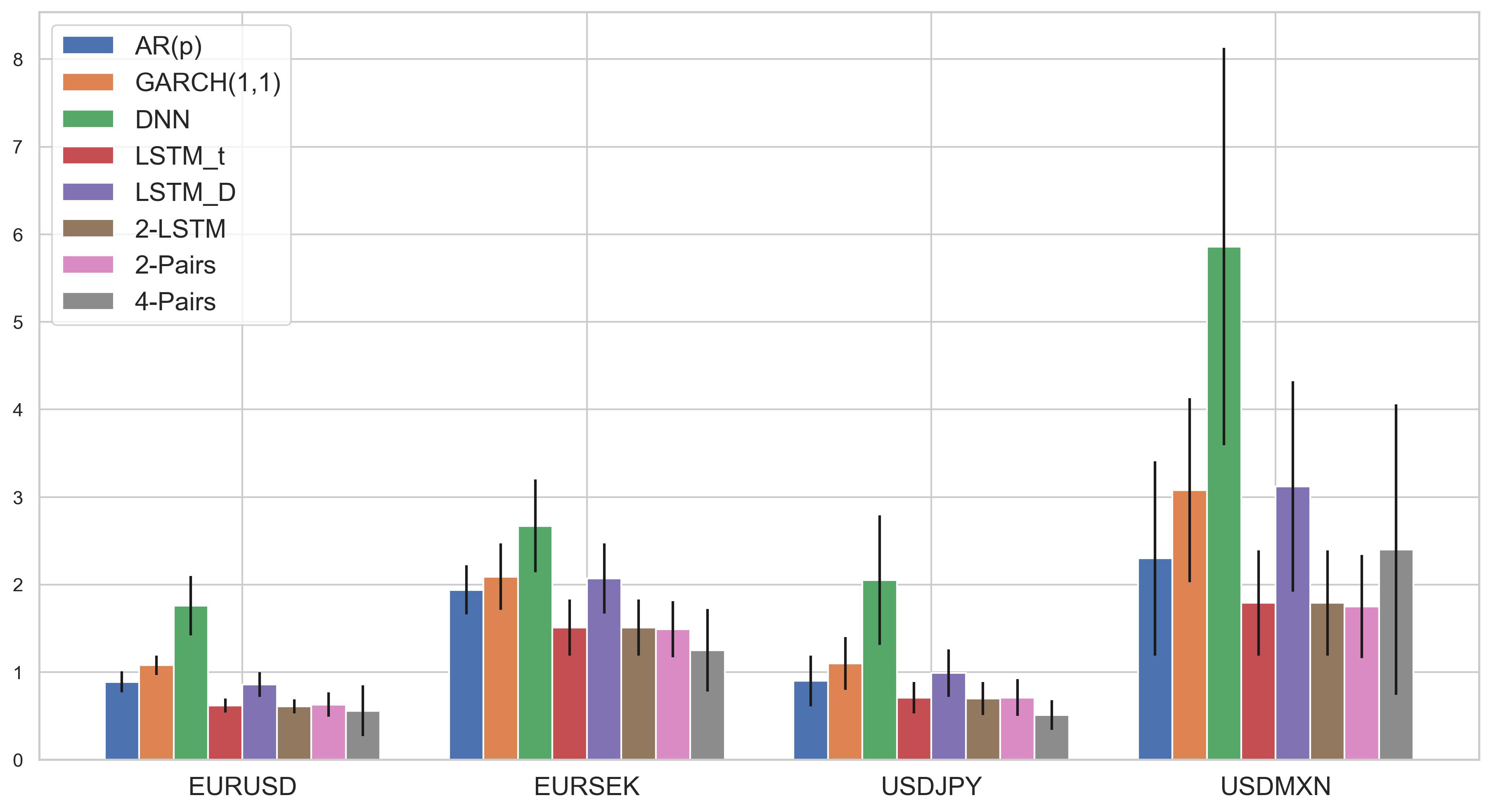}
        \caption{3-fold MSE ($\times 10^{-8}$) of daily log return comparison on testing sets. For 2-pairs-learning, we pair EURUSD (USDMXN) with EURSEK (USDJPY) respectively. (Color)}
        \label{mse_barplot}
    \end{figure}

\begin{figure}
    \centering
    \includegraphics[width=1\textwidth]{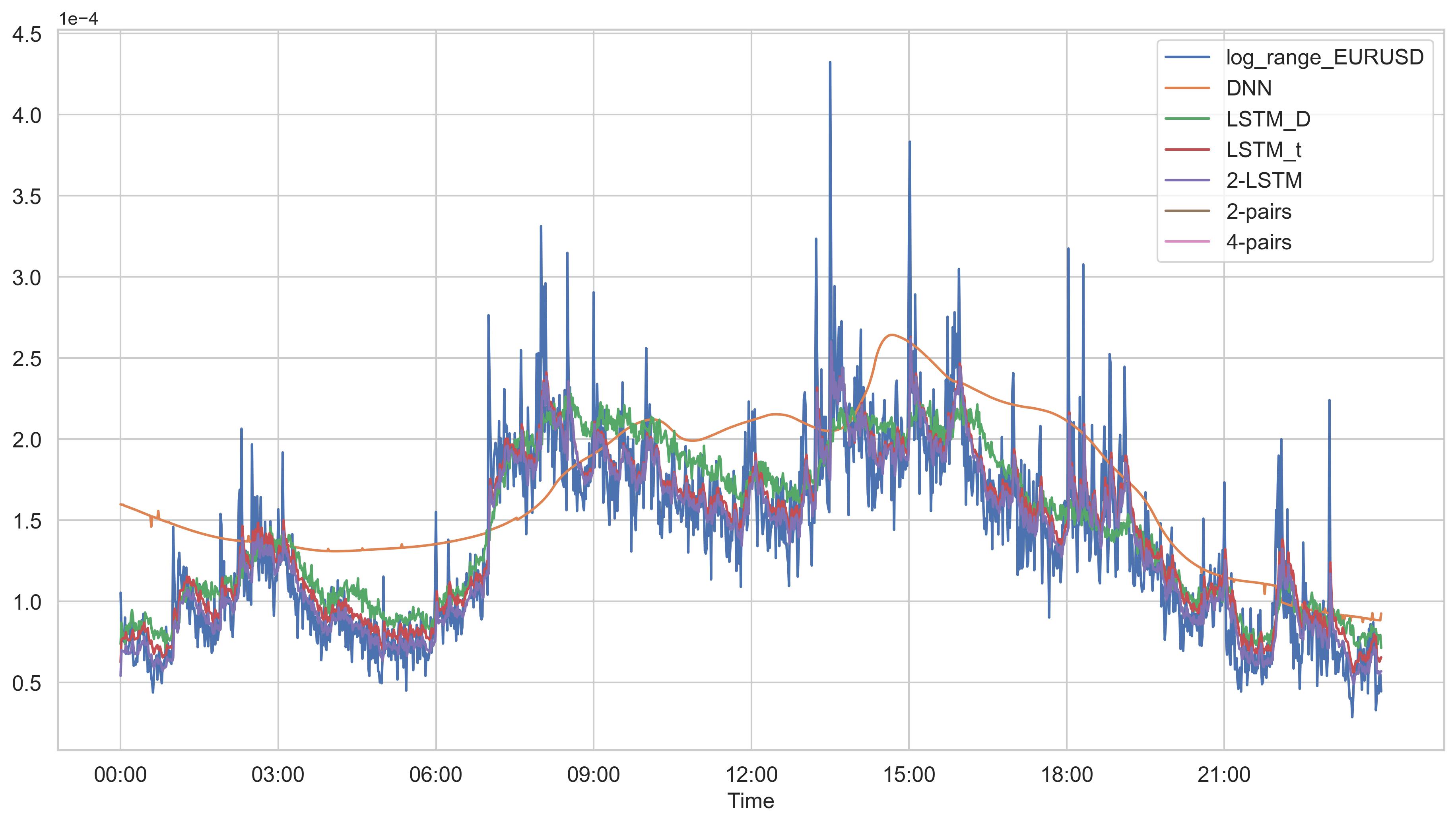}
    \caption{The comparison of predicted average of minutely range of different models on the testing data of EURUSD. (Color)}
    \label{test_plot}
\end{figure}

 Figure~\ref{mse_barplot} demonstrates that the proposed $p$-Pairs-learning 2-LSTM method consistently beats the other models in terms of the MSE on the testing sets. Except for the plain DNN, the neural network based model outperform the two traditional baselines. The LSTM based methods reduce the MSE noticeably from that of the plain DNN by incorporating the historical volatility information to capture the volatility clustering and seasonality in an effective way. In Table~\ref{mse_comp}, for example, for EURUSD, the 4-Pairs-learning 2-LSTM reduces the MSE by 37.07\% from the AR(3) model, 48.15\% from the GARCH(1,1) model and 68.18\% from the plain DNN. In Figure~\ref{test_plot}, of the fitting performance of the predicted average minutely range illustrates that the intra-day volatility is a useful input factor to boost the fitting performance and the models with this factor outperform the others (i.e. the plain DNN and LSTM$_D$).

 \begin{table}[]
    \centering
    \scalebox{0.8}{\begin{tabular}{l|>{\raggedleft}p{0.1\textwidth}|>{\raggedleft}p{0.1\textwidth}|>{\raggedleft}p{0.1\textwidth}|>{\raggedleft}p{0.1\textwidth}|>{\raggedleft}p{0.1\textwidth}|>{\raggedleft}p{0.1\textwidth}|>{\raggedleft\arraybackslash}p{0.1\textwidth}}
    \hline
    \multicolumn{8}{c}{\textbf{{\fontsize{10}{10}\selectfont(a) EURUSD}}}\\
    \hline
         &GARCH-(1,1)&Plain DNN & LSTM$_D$ & LSTM$_t$ & 2-LSTM & 2-Pairs-learning& 4-Pairs-learning \\
         \hline
         AR(3)& \textbf{-2.21}&  \textbf{-5.31}& {1.43} & \textbf{8.37} & \textbf{7.44}&\textbf{7.58}& \textbf{7.55}\\
         GARCH(1,1)& &  \textbf{-10.97}& 0.56 & \textbf{8.39} & \textbf{8.51}&\textbf{8.24}& \textbf{8.06}\\
        Plain DNN& & & \textbf{11.41}& \textbf{15.30} & \textbf{15.29} & \textbf{15.30}&\textbf{15.24}\\
        LSTM$_D$& & & &  \textbf{8.62}&  \textbf{8.77}&  \textbf{8.47}&
         \textbf{8.15}\\
         LSTM$_t$& & & & & -1.57&  -0.96&
        \textbf{-1.96}\\
         2-LSTM& & & & & & 0.26&
        0.05\\
         2-Pairs-learning& & & & & & & -0.90\\
         \hline

    \multicolumn{8}{c}{\textbf{{\fontsize{10}{10}\selectfont(b) EURSEK}}}\\
    \hline
         &GARCH-(1,1)&Plain DNN& LSTM$_D$ & LSTM$_t$ & 2-LSTM & 2-Pairs-learning& 4-Pairs-learning \\
         \hline
         AR(1)& \textbf{-7.62}&  \textbf{-9.69}& \textbf{2.64} & \textbf{10.57} & \textbf{10.14}&\textbf{9.32}& \textbf{9.11}\\
         GARCH(1,1) &&  \textbf{-8.98}& {1.90} & \textbf{9.57} & \textbf{9.59}&\textbf{9.43}& \textbf{9.39}\\
        Plain DNN && & \textbf{8.61}& \textbf{15.95} & \textbf{15.98} & \textbf{16.06}&\textbf{15.87}\\
        LSTM$_D$ && & & \textbf{10.00}&  \textbf{10.03}&  \textbf{9.88}&
         \textbf{9.64}\\
         LSTM$_t$ && & & & \textbf{5.96}&  \textbf{3.37}&
        {1.69}\\
         2-LSTM& && & & &\textbf{2.89}&
        1.36\\
         2-Pairs-learning& & & & & & &1.49\\
         \hline

    \multicolumn{8}{c}{\textbf{{\fontsize{10}{10}\selectfont(c) USDJPY}}}\\
    \hline
          &GARCH-(1,1)&Plain DNN& LSTM$_D$ & LSTM$_t$ & 2-LSTM & 2-Pairs-learning& 4-Pairs-learning \\
         \hline
         AR(2)& \textbf{-2.51}&  {1.64}& \textbf{9.82} & \textbf{8.87} & \textbf{8.91}&\textbf{9.44}& \textbf{9.74}\\
         GARCH(1,1)& &  \textbf{-11.67}& \textbf{-2.17} & \textbf{9.79} & \textbf{9.66}&\textbf{9.63}& \textbf{9.52}\\
        Plain DNN & & & \textbf{10.83}& \textbf{15.83} & \textbf{15.91} & \textbf{15.85}&\textbf{15.81}\\
        LSTM$_D$& & & & \textbf{7.56}&  \textbf{7.63}&  \textbf{7.62}&
         \textbf{7.35}\\
         LSTM$_t$& & & & & \textbf{6.41}&  {-1.69}&
        1.26\\
         2-LSTM& & & & & &\textbf{-3.23}&
        -0.56\\
         2-Pairs-learning& & & & & & & \textbf{2.70}\\
         \hline

    \multicolumn{8}{c}{\textbf{{\fontsize{10}{10}\selectfont(d) USDMXN}}}\\
    \hline
         &GARCH-(1,1)&Plain DNN& LSTM$_D$ & LSTM$_t$ & 2-LSTM & 2-Pairs-learning& 4-Pairs-learning \\
         \hline
         AR(5)& \textbf{-11.03}&  \textbf{-5.54}& \textbf{4.79} & \textbf{4.82} & \textbf{4.99}&\textbf{5.78}& \textbf{5.70}\\
         GARCH(1,1)& &  \textbf{-15.60}& \textbf{3.29} & \textbf{6.74} & \textbf{6.32}&\textbf{6.28}& \textbf{6.19}\\
        Plain DNN& & &\textbf{17.25}& \textbf{20.36} & \textbf{20.35} & \textbf{20.34}&\textbf{20.43}\\
        LSTM$_D$& & & & \textbf{9.36}&  \textbf{9.37}&  \textbf{9.02}&
         \textbf{8.60}\\
         LSTM$_t$& & & & &\textbf{-3.26}&  \textbf{-3.83}&
        0.42\\
         2-LSTM& & & & & &\textbf{-3.20}&
        1.33\\
         2-Pairs-learning& & & & & & &\textbf{-3.88}\\
         \hline
    \end{tabular}}
    \caption{The comparison table of the pairwise Diebold-Mariano test statistics among models for each currency pair. Positive numbers indicate that the column model outperforms the row model.
Bold font indicates that the difference is significant at 5\% level or better for individual tests.}
    \label{dm_test}
\end{table}

 In Tabel~\ref{dm_test} of the pair-wised DM test results, we observe that for EURUSD, the 2-LSTM and 2-Pairs-learning outperform other methods and have statistically non-significant difference between them. For EURSEK, the best model is 4-Pairs-learning, which has notably better performance against others. For USDJPY, the optimal method is 2-LSTM, which have all positive DM statistic against other models. For USDMXN, LSTM$_t$ and 4-Pairs-learning have statistically similar performance, while outperform others. In a summary,
 the LSTM based models, especially LSTM$_t$ and $p$-Pairs-learning have the superior performance over others based on the statistical test.

%

\subsection{Comparison between LSTM based models}
Since LSTM$_t$, LSTM$_D$, $2$-LSTM models and $p$-Pairs learning models all use historical volatility information and the recurrent structure in dealing with time series data, we compare their forecasting ability in this section. As shown in Table~\ref{mse_comp}, the 2-LSTM has a significant improvement over the LSTM$_D$ model, while it does not increase performance significantly over the LSTM$_t$ model. Since the LSTM$_t$ uses only the previous minutely information and the LSTM$_D$ use solely the volumes of previous $p_d$ days at the same minute, from the results we can conclude that the LSTM has better performance when using the intra-day historical data. That 2-LSTM improves little performance over the LSTM$_t$ can be justified by the stronger auto-correlation of intra-day volatility than that of inter-day volatility. Figure~\ref{acf_plot} shows that the short-term intra-day historical data has auto-correlation around 0.5 whereas the highest auto-correlation of inter-day data can only reach 0.3.

For 2-Pairs learning, each currency pair includes one liquid currency (i.e. EURUSD and USDJPY) and the other currency with poor liquidity (i.e. EURSEK and USDMXN). From Table~\ref{mse_comp}, we can see the prediction results are slightly improved for the less liquid currency (MSE reduced by 1.32\% for EURSEK and 2.23\% for USDMXN comparing with $2$-LSTM method), as more useful information is incorporated by coupling with the liquid one.

For 4-Pairs learning, except for the USDMXN, all the other results are significantly boosted compared with 2-pairs learning considering the average of MSE (the average of MSE reduced by 11.11\% for EURUSD, 16.1\% for EURSEK and 28.17\% for USDJPY). The results suggest that although some of Forex pairs do not have same base or quote currencies, they may share same market information. An interesting and important question for future investigation is that how to choose the pairs of currencies such that the best results can be achieved.

\begin{table}[!ht]
     \centering
     \begin{tabular}{r|c|c|c|c}
     \toprule
          & EURUSD& EURSEK& USDJPY & USDMXN \\
          \midrule
       AR(p) & 0.89 $\pm$ 0.12 & 1.94 $\pm$ 0.28 &0.90 $\pm$ 0.29& 2.30 $\pm$ 1.11 \\

       GARCH(1,1) & 1.08 $\pm$ 0.11  & 2.09 $\pm$ 0.38 & 1.10 $\pm$ 0.30 & 3.08 $\pm$ 1.05 \\

       Plain DNN & 1.76 $\pm$ 0.34 & 2.67 $\pm$ 0.53&2.05 $\pm$ 0.74& 5.86 $\pm$ 2.27  \\

           LSTM$_t$ & 0.62 $\pm$ 0.08& 1.51 $\pm$ 0.32&0.71 $\pm$ 0.18 & 1.79 $\pm$ 0.60\\

       LSTM$_D$ & 0.86 $\pm$ 0.14& 2.07 $\pm$ 0.40& 0.99 $\pm$ 0.27 & 3.12 $\pm$ 1.20\\

       2-LSTM & {0.61} $\pm$ {0.08} & 1.51 $\pm$ 0.32& {0.70} $\pm$ {0.19} & 1.79 $\pm$ 0.60\\

       2-Pairs-learning  & 0.63 $\pm$ 0.14 & {1.49} $\pm$ {0.32} & 0.71 $\pm$ 0.21 & \textbf{1.75} $\pm$ \textbf{0.59} \\

        4-Pairs-learning  & \textbf{0.56} $\pm$ \textbf{0.29} & \textbf{1.25} $\pm$ \textbf{0.47} & \textbf{0.51} $\pm$ \textbf{0.17} & {2.40} $\pm$ {1.66} \\
        \bottomrule
     \end{tabular}
     \caption{3-fold MSE ($\times 10^{-8}$) of daily log return comparison of all methods for different currency pairs on the testing sets. For pairs-learning, we pair EURUSD (USDMXN) with EURSEK (USDJPY) respectively.}
     \label{mse_comp}
 \end{table}

\subsection{Sensitivity analysis}
\begin{figure}[h]
    \centering
    \includegraphics[width= 0.7\textwidth]{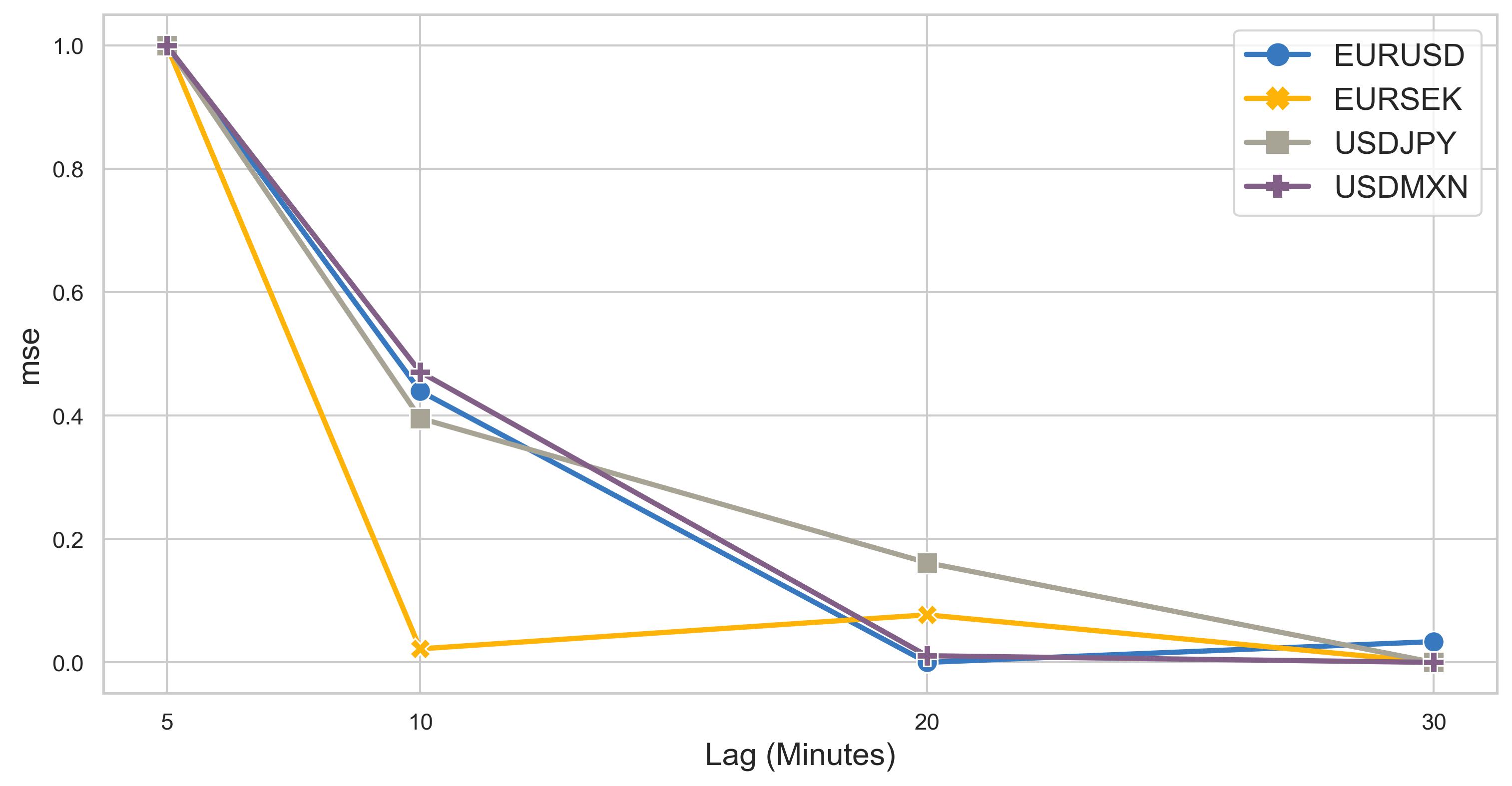}
    \caption{The sensitivity analysis of 4-Pairs-learning w.r.t. the lag value $p$. (Color) }
    \label{sensi_analyse}
\end{figure}

We study the sensitivity of our best model, 4-Pairs-learning 2-LSTM, in this subsection. In Figure~\ref{sensi_analyse}, we min-max normalise the MSE of each currency pair separately and show the sensitivity analysis of 4-pairs-learning w.r.t. the lag value $p$ for 2-LSTM. It displays that when $p$ increases, the MSE of prediction on testing data decreases and it becomes steady when $p$ is larger than 20.


\section{Conclusion}\label{Section_Conclusion}
In this work, we consider the volatility forecasting problem in Forex market. With the divide-and-conquer approach driven by domain knowledge of the market, we develop deep learning based models to provide end-to-end solutions to the forecasting challenge and capture the patterns in the intra-day volatility. In Table~\ref{models_summary}, we summarise the advantages and disadvantages of the proposed models. Our $p$-Pairs-learning 2-LSTM model consistently outperforms conventional methods and other deep learning models on all currency pairs in the comparison of MSE as well as the DM test. This work shows the strength of a deep learning model constructed carefully by the domain knowledge. The proposed method can be combined with volatility-targeted strategy to build an end-to-end trading system.
\newpage
\input{appendix}

\medskip

\bibliography{references.bib}

\end{document}

%% file: appendix.tex
\begin{appendix}
\section{Neural network preliminaries}\label{section_appendix}
\subsection{Supervised Learning Framework}\label{section_supervised_learning}
The supervised learning aims to learn a functional relationship between the input and output based on the observed input-output pairs. Specifically, let us consider the dataset composed with $N$ input-output pairs $\mathcal{D}:= \{(x_i,y_i)\}_{i = 1}^{N}$, where the input $x \in\mathbb{R}^d$ and the output $y \in\mathbb{R}^e$. Assume that there exists a continuous map $f: \mathbb{R}^{d} \rightarrow \mathbb{R}^{e}$, such that $\forall i \in \{1, \cdots, N\}$, 
\begin{eqnarray*}
y_{i} = f(x_{i}) + \varepsilon_{i},
\end{eqnarray*}
where $\varepsilon_{i}$ is a $\mathbb{R}^{e}$-valued random variable, and $\mathbb{E}[\varepsilon_{i} \vert x_{i}] = 0$. One may postulate $f$ by a parameterized model $f_{\theta}$, which is fully characterized by $\theta$. The objective of the supervised learning is to learn $f_{\theta}$ such that the model estimated output $\hat{y}_{i}$ as close to the actual output $y_{i}$ as possible. More precisely, the learning process can be formulated as to find the optimal parameter $\theta^{*}$ such as to minimize the loss function, denoted by $\mathcal{L}( \theta | \mathcal{D})$, which measures the prediction performance of the parameter $\theta$ for the given dataset $\mathcal{D}$. In formula,  
\begin{equation*}
    \theta^{*} = \mathop{\text{argmin}}_{\theta}\limits{}\mathcal{L}( \theta \vert \mathcal{D}).
\end{equation*}

\paragraph{Loss function}
There are different choices of loss functions. The popular loss function of the classification problem is the cross entropy. In contrast to it, the most commonly used loss function of the regression problem is the mean squared error (MSE), i.e.
\begin{eqnarray}
\mathcal{L}(\theta | \mathcal{D})=\frac{1}{N}\sum_{i=1}^N |f_{\theta}(x_i)-y_i|^2.
\label{mse}
\end{eqnarray}
where $|.|$ is the $l^{2}$ norm. MSE loss function is also the one we use for this paper.
\paragraph{Model} There are various candidates for the model $f_{\theta}$ under the supervised learning framework, such as support vector machine, linear regression, logistic regression, decision trees, neural network, etc., among which neural network is of our interest.
\subsection{Neural Network}
\subsubsection{Deep Neural Network}
Conventionally, the deep neural network (DNN) is a multi-layer neural network with $L>2$. Let $l\in\{1,\dots,L-1\}$ be the index for hidden layers and $n_l$ be the number of neurons, i.e. the number of elements in the output vector of the $l^{th}$ layer. Layers in DNN are defined in the following recursive way:
\begin{itemize}
    \item Input layer $h_0: \mathbb{R}^{d}\to \mathbb{R}^{d}$, \begin{equation*}
        h_0(x) = x,\hspace{2pt} \forall x \in \mathbb{R}^d
    \end{equation*}

    \item Hidden layer $h_l: \mathbb{R}^{d}\to \mathbb{R}^{n_l}$,
    $$h_l(x) =  \sigma_l({W_l}h_{l-1}(x) + b_l), \hspace{2pt} \forall x \in \mathbb{R}^d$$
    where $W_l\in \mathbb{R}^{n_l\times n_{l-1}}$ is the weight matrix, $b_l\in \mathbb{R}^{n_l}$ is the bias term and $\sigma_l$ is called \textbf{activation} function.
    \item Output layer $h_L: \mathbb{R}^{d}\to \mathbb{R}^{e}$, $$h_L(x) =  \sigma_L({W_L}h_{L-1}(x) + b_L), \hspace{2pt} \forall x \in \mathbb{R}^d$$
    where $W_L\in \mathbb{R}^{e \times n_{L-1}}$ is the weight matrix, $b_L\in \mathbb{R}^{e}$ is the bias term and $\sigma_L$ is activation function.
\end{itemize}
The parameter set is $\Theta := \{W_l, b_l\}_{l=1}^L$. For the DNN described above, the estimated conditional expectation of the output given any given input $x$ is $h_{L}(x)$. 

\subsubsection{Long Short-Term Memory}
The Long Short-Term Memory (LSTM) is designed to capture the long-term dependencies of time series data and cope with gradient vanishing and exploding~\cite{lstm_vocal_sak}. LSTM is also composed of the input layer, the hidden layer and the output layer. In the hidden layer at time step $t$, the memory cell $c_t$ works as the main part of LSTM. In the memory cell, there are three gates denoted as the input gate $i_t$, the output gate $o_t$ and the forget gate $f_t$. Let $x_t$ be the input vector, $h_t$ be the output of the hidden layer, then the computation in the memory cell is shown in the following equations:
\begin{eqnarray*}
f_t &=& \sigma_g(W_f x_t + U_f h_{t-1} + b_f),\\
i_t &=& \sigma_g(W_i x_t + U_i h_{t-1} + b_i),\\
o_t &=& \sigma_g(W_o x_t + U_o h_{t-1} + b_o),\\
\Tilde{c}_t &=& \sigma_h(W_c x_t + U_c h_{t-1} + b_c),\\
c_t &=& f_t\odot c_{t-1} + i_t\odot \Tilde{c}_t,\\
h_t &=& o_t\odot \sigma_h(c_t),
\end{eqnarray*}
where $\{W_f,W_i,W_o,W_c, U_f, U_i, U_o,U_f\}$ is the set of weight matrices, $\{b_f,b_i,b_o,b_c\}$ is the set of bias terms. $\odot$ is the elementise muliplication of two vectors. $\sigma_g$ is the sigmoid function and $\sigma_h$ is the hyperbolic tangent function. We denote the LSTM model as $R((x_t)_t)$. Here $\Theta:=\{U,W,b\}$ is the parameter set. For the ease of notation, we denote the LSTM model by the function $R: \mathbb{R}^{d \times T} \rightarrow \mathbb{R}^{e}$: 
\begin{eqnarray*}
(X_{t})_{t = 1}^{T}\mapsto h_{T}.
\end{eqnarray*}

\section{Hyper-parameter tuning details}\label{hp_tune}
\begin{figure}
    \centering
    \includegraphics[width= 1\textwidth]{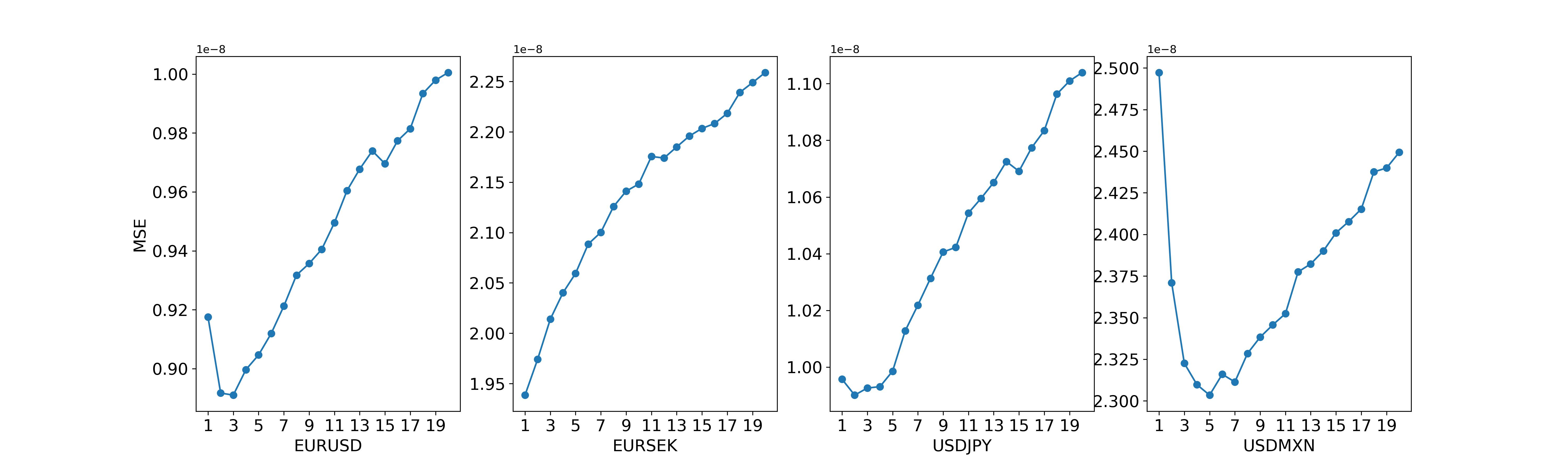}
    \caption{The MSE of AR($p$) models for different currency pairs with respect to lag order $p$.}
    \label{ar_order_select}
\end{figure}

We tune the lag order $p$ of AR model for all currencies pairs. The results are shown in Figure~\ref{ar_order_select}, which shows the corresponding optimal lag orders.

We conduct the hyper-parameter tuning for the proposed models on the EURUSD dataset. We apply the optimal hyper-parameters for the other FX currency pairs. The details are as follows:
\begin{itemize}
    \item Plain DNN: We implement grid search for number of the hidden layers $L\in \{2,4,6,8,10\}$ and the number of hidden neurons $n \in \{5,10,20,30\}$ of each layer. The results are shown in Table~\ref{hp_dnn}. 
    \item Plain LSTM, $2$-LSTM and $p$-pairs learning: We implement parameter tuning for lag value $p:=p_t=p_d \in \{5, 10, 20, 30\}$. Results are shown in Table~\ref{hp_rnn}.
\end{itemize}

\begin{table}[htbp]
  \centering
  
    \begin{tabular}{cccccc}
    \toprule
    &\multicolumn{5}{c}{L}\\
    \cmidrule(lr){2-6}
       N   & 2 & 4 & 6 & 8 & 10 \\
          \cmidrule(lr){1-1}
\cmidrule(lr){2-6}
    5     & 2.75$\pm$0.56 & 2.45$\pm$0.58 & 2.33$\pm$0.62 & 2.34$\pm$0.41 & 2.26$\pm$0.30 \\
    10    & 2.54$\pm$0.52 & 2.12$\pm$0.47 & 1.96$\pm$0.58 & 2.01$\pm$0.56 & 1.91$\pm$0.42 \\
    20    & 2.27$\pm$0.71 & 2.08$\pm$0.28 & 1.84$\pm$0.39 & 1.88$\pm$0.50 & 1.89$\pm$0.30 \\
    30    & 2.01$\pm$0.62 & 1.85$\pm$0.47 & 1.76$\pm$0.34 & 1.78$\pm$0.39 & 1.76$\pm$0.31 \\
    \bottomrule
    \end{tabular}%
    \caption{Number of layers and hidden neurons tuning of the plain DNN model.}
  \label{hp_dnn}%
\end{table}%

\begin{table}[htbp]
  \centering
    \begin{tabular}{rcccc}
    \toprule
    &\multicolumn{4}{c}{$p$}\\
    \cmidrule(lr){2-5}
        Models & 5 & 10 & 20 & 30 \\
          \cmidrule(lr){1-1}
\cmidrule(lr){2-5}
    LSTM$_t$ & 0.635$\pm$0.08 & 0.629$\pm$0.09 & 0.621$\pm$0.08 & \textbf{0.619$\pm$0.08} \\
    LSTM$_D$ & 0.921$\pm$0.32 & 0.894$\pm$0.27 & \textbf{0.863$\pm$0.14} & 0.872$\pm$0.19 \\
    2-LSTM & 0.637$\pm$0.10 & 0.625$\pm$0.08 & 0.613$\pm$0.08 & \textbf{0.611$\pm$0.08} \\
    2-pairs-learning  & 0.642$\pm$0.16 & 0.637$\pm$0.14 & \textbf{0.630$\pm$0.14} & 0.633$\pm$0.15 \\
    4-pairs-learning  & 0.580$\pm$0.29 & 0.573$\pm$0.29 & \textbf{0.564$\pm$0.29} & 0.567$\pm$0.29 \\
    \bottomrule
    \end{tabular}%
    \caption{$p$-lag value tuning of the LSTM based methods.}
  \label{hp_rnn}%
\end{table}%

\end{appendix}

\clearpage